\documentclass[twocolumn,showpacs,amsmath,amssymb,aps,prb]{revtex4}

\usepackage{graphicx} % Include figure files

\usepackage{bm} % bold math
\usepackage{epsfig}

\begin{document}

\title{Heterogeneous dynamics of the three dimensional Coulomb glass
out of equilibrium}
\author{Alejandro B. Kolton}
\affiliation{Universit\'e de Gen\`eve, DPMC, 24 Quai Ernest Ansermet,
CH-1211 Geneve 4, Switzerland}
\author{D. R. Grempel}
\affiliation{CEA-Saclay/DSM/DRECAM/SPCSI, 91191 Gif-sur-Yvette CEDEX, France}
\author{Daniel Dom\'{\i}nguez}
\affiliation{Centro At\'{o}mico Bariloche,
 8400 San Carlos de Bariloche, R\'{\i}o Negro, Argentina.}

\today

\begin{abstract}
The non-equilibrium
 relaxational properties of a three dimensional Coulomb glass model
 are investigated by 
 kinetic Monte Carlo simulations.
Our results suggest a transition from stationary to non-stationary 
dynamics at the equilibrium glass transition temperature of the
system.  Below the transition the 
dynamic correlation functions loose time translation invariance and electron 
diffusion is anomalous. 
Two groups of carriers can be identified at each time scale, electrons
whose motion is diffusive within a selected time window and electrons
that  during the same
 time interval remain confined in small regions in space.
During the
relaxation that follows a temperature quench
an exchange of electrons between these two groups takes place
and the non-equilibrium excess of diffusive electrons 
 initially present decreases logarithmically with time 
as the system relaxes.
This 
bimodal dynamical heterogeneity persists 
at higher temperatures when time translation 
invariance is restored and electron diffusion is normal.
 The occupancy of the 
two dynamical modes is then stationary  and its temperature 
dependence reflects a crossover between a low-temperature 
regime with a high concentration of electrons forming 
fluctuating dipoles and a high-temperature regime in which the 
concentration of diffusive electrons is high. 
\end{abstract}


\maketitle

\newpage
\section{INTRODUCTION}

Recent experimental studies of hopping conductance in Anderson 
insulators showed striking non-equilibrium effects  
that persist for long times at low temperature.
~\cite{films:experiments,films:goldman,films:aging,films:ee-int,orlyanchik}
These results support 
early theoretical predictions of a low-temperature 
glassy phase in interacting disordered electron systems in the strongly  
localized limit. \cite{review:coulombglass1,coulomb-glass} 
In this regime
the essential physics is well 
captured by the Coulomb glass model that  
describes a system of interacting 
electrons hopping between randomly distributed sites that correspond  
to the localization centers of the single-electron 
wavefunctions. 
\cite{review:coulombglass1,coulomb-glass,pollak1,davies,lee,yu1} 

Although Coulomb glass models have  
been extensively studied during 
the last thirty years and their equilibrium properties 
are by now fairly well  understood
\cite{review:coulombglass1,coulomb-glass,pollak1,davies,lee,yu1,newstuff}
 much less is known about the properties of
 Coulomb glasses out of equilibrium.
\cite{ortuno,yu2,tsigankov,grempel} 

In Reference \onlinecite{grempel} the off-equilibrium dynamics of the Coulomb 
glass was investigated by studying the     
scaling properties of the non-stationary correlation and 
response functions, a tool often used
 in the study of other glassy systems such as structural and
 spin glasses. \cite{review:glasses1,review:glasses2,review:glasses3,CuKu} . 
It has recently been realized that further insights on the nature of 
the glassy phase can be gained from the analysis
 of  dynamical heterogeneities. \cite{dynhet,dynhet2,ritort} 
This approach is particularly appealing in the context of Coulomb  
 glasses since 
 the low-temperature  
hopping conductance is dominated by the diffusion of carriers
 on a set of conducting percolation paths. 
\cite{review:coulombglass1,coulomb-glass} 
We expect interactions between the  
electrons to have strong effects on this type of motion  
since hops of individual carriers  
alter the effective random potential 
felt by the others and thus modify the structure of the percolating network. 
Fluctuation effects of this type were shown to lead to  
 low-frequency $1/f$-like noise in the
 conductance of Coulomb glasses. \cite{noise1,noise2}

In this paper we investigate the dynamical properties of the  three
dimensional  random-site Coulomb glass model by  kinetic Monte Carlo
simulation using a realistic  microscopic dynamics that favors the
emergence of local effective  constraints in the kinetics. In our
simulations  the system is initially quenched from infinite temperature
to  a working temperature $T$ and its evolution in time is
characterized for different values of $T$ through the time dependence
of the relevant correlation functions. 

One of our main observations is the appearance  of a dynamical 
crossover 
from equilibrium dynamics  to slow non-equilibrium dynamics
at a temperature $T_g\sim T_c$, where $T_c$ is   
the equilibrium freezing transition temperature of
the model. \cite{yu1} This crossover takes place even for  relatively
small system sizes and occurs at the temperature at which the 
equilibration time of the finite sample  becomes much longer than the
time-scale of the simulation. In  this regime the time-dependent
correlation functions  exhibit slow relaxation and have aging
properties.  We found that the dynamics of the Coulomb glass is
heterogeneous as observed  in other glassy systems.
\cite{dynhet,dynhet2} Heterogeneities can be characterized by
examination of the    evolution of diffusion fronts. A statistical
analysis of the electron  trajectories over a fixed time window shows
that most carriers belong to one or the other of two groups. The first group
 is that of those electrons that diffused away from their initial position 
during the chosen time interval. The second group
 is that of the electrons that
remained confined in relatively small regions in space during that time.  
An exchange of
electrons between these two dynamical modes takes place  as the
system relaxes after a temperature quench.    In the aging regime this
exchange is very slow and the non-equilibrium excess of electrons  with
metallic hopping present in the system right after the  quench
decreases logarithmically with time. In this temperature range the
diffusion is anomalous.  We found that the mean squared displacement
 $\langle \Delta x^2(t)\rangle \sim t^\eta$  where the 
exponent is $\eta <1 $ and depends on the age of the system. 
In the equilibrium regime time translation 
invariance is recovered and we observe normal diffusion. However,
 the bimodal dynamical heterogeneity still persists. The occupancy of the 
two dynamical modes is stationary  and its temperature 
dependence reflects a crossover from a low-temperature 
regime with a high concentration of electrons forming 
fluctuating dipoles to a high-temperature regime in which the 
concentration of diffusive electrons is high. 
 
The paper is organized as follows.  Section \ref{model}   contains a
description of the model and of our numerical method. In  Section
\ref{charge} we discuss the properties of the local-density
autocorrelation function in and out of equilibrium. Section
\ref{current} is devoted to the analysis of electron diffusion in the
system. In Section  \ref{hetero} we study the statistical  properties
of the diffusion fronts and show that they provide evidence for the
existence of heterogeneous transport in the system.  Finally, we
summarize  the conclusions of our study in  Section \ref{conclusions}.

\section{The model}
\label{model}
The Hamiltonian of the classical three dimensional Coulomb glass 
 is~\cite{coulomb-glass} 
\begin{equation}\label{hamil}
H=\sum_i n_i \varphi_i + {e^2 \over 2 \kappa} \sum_{i\ne j} {(n_i
- K) (n_j - K) \over \left|{\bf R}_i - {\bf R}_j\right|}\;,
\end{equation}
where ${\bf R}_i$ denotes center of localization of a single-particle 
localized electronic 
state, $\varphi_i$ the  energy of the state 
and  $e$ and $\kappa$ 
are the electron charge and the medium's  dielectric constant, 
respectively. Strong on-site correlations limit the occupancy of the
electronic states to  $n_i=0,1$. Charge neutrality is assured by a
uniform compensating  positive charge density $K=(1/N)\sum_i n_i$.

The positions ${\bf R_i}$ of the localized states and their energies
$\varphi_i$ are both random variables. However, it is common practice
to study two complementary simplified versions of the model. These are
respectively   referred to as the lattice model and the random site
model in Reference \onlinecite{tsigankov}. In the lattice model the
sites are assumed to lie on a regular cubic lattice and only the
randomness in the energies $\varphi_i$  is taken into consideration.
Conversely, in the random site  model~\cite{lee,yu1} only the
positional disorder is taken into account and $\varphi_i=0$. It is
standard practice to concentrate on the  particle-hole  symmetric case
$K=1/2$ for which the analysis of the results is simpler. 
 
While it has been established that the three dimensional random-site
model  has an equilibrium glass transition at low
temperature~\cite{yu1}  it is not yet clear whether this is also the
case for the lattice model. \cite{transiciondeglass,newstuff} 
Therefore, we shall only discuss the dynamics of the random-site model
in this paper.  

To model the dynamics of the  system we let it evolve through
sequential single-electron hops from occupied sites $a$ to empty
sites $b$. The transition rate, that mimics phonon assisted processes, is
\begin{eqnarray}\label{rate}
\Gamma_{a\rightarrow b}& =& \tau_0^{-1} e^{-{2 R_{ab}
/ \xi}}\;  
\min\left[1, e^{-\Delta E_{ab}/T} \right]\;,
\end{eqnarray}
where $\tau_0$ is a microscopic time, $\xi$ is the 
spatial extension of the localized wavefunctions and 
$R_{ab} \equiv \left|{\bf R}_a - {\bf R}_b\right|$. 
$\Delta E_{ab}$, the total energy
difference  in the transition, is given by 
\begin{eqnarray}
\label{energies} \Delta E_{ab}& =& \epsilon_b - \epsilon_a - {e^2
\over \kappa R_{ab}}\;,\;\;\;\epsilon_a = {e^2 \over \kappa}
\sum_{b\ne a} {\delta n_b  \over R_{ab}}\;.
\end{eqnarray}
$\delta n_i \equiv n_i - K$.

The first factor in Eq.(\ref{rate})  reflects
 the exponential decay of the electron-phonon matrix element between  
two electronic wavefunctions centered at positions ${\bf R}_a$
 and ${\bf R}_b$. The second factor is the thermal part of the 
transition probability. 
In Monte Carlo simulations performed by other authors 
 the transition probability is taken independent of the distance $R_{ab}$ 
[the first exponential in the equivalent of Eq.(2) is absent]. 
\cite{davies,yu1} This type of non-local dynamics,  
convenient for rapid equilibration, may not be appropriate   
for the study of off-equilibrium relaxation. 
With the local dynamics of Eq.~\ref{rate} 
 electron hops that {\it decrease} the energy are essentially 
restricted to a region whose linear size is the 
 localization length $\xi$. This introduces dynamic constraints that
contribute to make the relaxation out of an excited configuration
slower.    

In our simulations we take  the mean distance between sites $a_0$ as
the unit of length, the Coulomb energy $E_C = e^2/(\kappa a_0)$ as the
unit of energy and  we choose for convenience  $ \xi =a_0 $.

We simulated systems of $N=L^3$ sites and $M=N/2$ electrons for samples with 
$L=$6, 8 and 10 in  the temperature range $0.01 \le T\le 0.1$. 
The localization centers are distributed randomly  
and uniformly inside a computational cubic box of side $L$ 
and we take periodic boundary conditions in all directions.  
To simulate a quench from high temperature we start from a 
random electron configuration 
at $t=0$ and let the system freely evolve with the  
dynamics~(\ref{rate})  at the working temperature $T$. 
The elementary Monte Carlo move consists of an attempt to move 
 an electron from a randomly chosen occupied site $a$ to an 
empty site $b$.  Once $a$ is chosen, the destination site 
$b$ is chosen randomly using the 
probability distribution of the hoppings, 
$P\left(R_{ab}\right)\propto \exp\left(-2 R_{ab}\right)$.
A cutoff at $R_{ab}=L/2$ is imposed by our using of 
periodic boundary conditions and restricts us to a range of   
temperatures for which hops at distances  $R_{ab} \sim L$ can be neglected. 
The
probability of acceptance of the move is the thermal 
factor in Eq.~(\ref{rate}). If the hop is accepted, the vector 
displacement of 
the hopping electron ${\bf \delta r}$ is defined as the vector going 
 from site $a$ to the closest periodic image of site $b$. 
A Monte Carlo step (MCS)
consists of $N$ hopping attempts. Our runs were 
 typically $2\times 10^6$ MCS long. Physical quantities were
monitored as a function of time for each sample and the results were 
averaged over between 150 and 600 realizations of the disorder and 
initial conditions, depending on the system size and temperature.

\section{The local density autocorrelation function}
\label{charge}
The two-time charge autocorrelation function is \cite{grempel}
\begin{equation}
C(t + t_w,t_w) = {2 \over M} \sum_i \langle \delta n_i(t+ t_w)\;
\delta n_i(t_w)\rangle\;, \label{C}
\end{equation}
where the brackets 
denote the average over configurational
disorder, initial conditions, and thermal noise, and the  
waiting time $t_w$ is the time elapsed since the quench from 
infinite temperature. 

The function $C(t + t_w,t_w)$ is the overlap 
of the charge configurations at times $t+t_w$ and $t_w$. If 
$t_w$  is larger than the
equilibration time $\tau_{\rm eq}$ the state of the system is
time-translational invariant and the correlation function depends
only on the time difference $t$. Otherwise, $C$ depends on both
$t$ and $t_w$.

\begin{figure}[tbp]
\centerline{\epsfxsize=\hsize \epsfbox{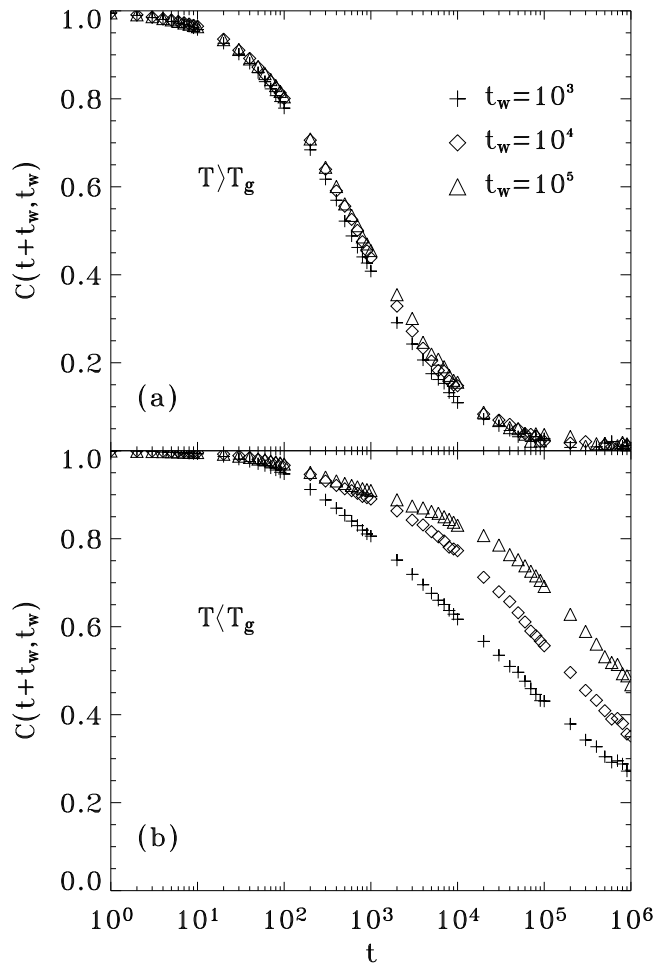}}
\caption{Two-time charge autocorrelation functions for $T=0.07$ (a), 
and $T=0.03$ (b). Aging effects (loss of time translation invariance) 
become appreciable bellow the crossover temperature 
$T_g \approx 0.05$.}
\label{fig:fig1}
\end{figure}

We describe in the following results for a system of linear 
size $L=10$. Fig.~\ref{fig:fig1} displays   
$C(t + t_w,t_w)$ {\it vs.} $t$ for three waiting times,  
$t_w=10^3, 10^4$ and $10^5$ and two representative cases,  
 $T=0.07$ [Fig.\ref{fig:fig1}(a)]  and $T=0.03$ 
[Fig.\ref{fig:fig1}(b)]. 
 \noindent
We observe that in the first case $C(t+t_w,t_w) \approx C(t)$, {\it
i.e.}  the system is time translational invariant. This means that
equilibrium was reached  within a time shorter than the shortest of the
waiting times considered, $\tau_{eq}(T)<10^3$.   $C(t)$ is thus the
equilibrium relaxation  function. In the second case, however, time
translation invariance is lost and $C(t+t_w,t_w)$ depends on both time
arguments, a phenomenon known as aging.  The system was thus unable to 
equilibrate within the time scale of the  simulation.  Note that for
each value of $t_w$ the relaxation is very slow (roughly  logarithmic)
and becomes slower with  increasing $t_w$. 

We found that non-equilibrium relaxation appears  below a dynamic crossover
temperature $T_g \sim 0.05$.  The value of $T_g$ is remarkably close to
the equilibrium transition temperature of the random-site model $T_c =
0.043$ determined in Ref.~[\onlinecite{yu1}].

Further insights on the properties of the correlation functions can be
gained by performing a scaling analysis of the data. We discuss
separately the cases of high and low temperatures.

\subsection{$T>T_g$}

\begin{figure}[tbp]
\centerline{\epsfxsize=\hsize \epsfbox{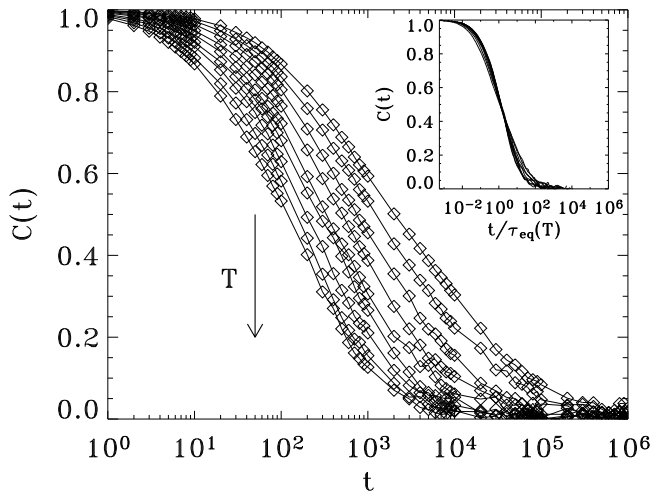}}
\caption{The charge 
autocorrelation function at equilibrium for  $T>T_g$. The curves correspond to 
$T$=0.10, 0.095, 0.09, 0.085, 
0.080, 0.075, 0.070, 0.065, 0.060, 0.055 and 0.050, from bottom to top. 
The inset shows the same data but 
represented as a function of the rescaled time $t/\tau_{eq}(T)$,
 with $\tau_{eq}(T)=\exp(0.45/T)$.
}
\label{fig:fig2}
\end{figure}

In this temperature region the system reaches equilibrium within the
simulation time. We display in Fig. \ref{fig:fig2} the equilibrium
correlation function  $C(t)$ obtained for several temperatures    in
the range $0.05\le T\le 0.1$.  
As shown in the inset to Fig.~\ref{fig:fig2}, the curves for the
various temperatures collapse rather well into a single master curve
when $C(t)$ is  represented as a function of the scaled variable
$t/\tau_{eq}(T)$, where $\tau_{eq}(T) = \exp(T_0/T)$ with  $T_0 \sim
0.45$. The equilibrium relaxation time thus obeys an  Arrhenius law
above the dynamic crossover temperature. Qualitatively similar results
were obtained in  simulations of the two-dimensional version of the
random-site model.\cite{lee,grempel} In previous work on the 3D model
using the non-local dynamics described above a  power-law divergence of
the relaxation time was reported at the   transition temperature
$T_c$.\cite{yu1} In our case this temperature $T_c$ is of the same order as
the dynamic crossover temperature $T_g$ for which our samples stay  
out of equilibrium during the entire
simulation time. Therefore, we have no access to the equilibrium
critical dynamics of the model.

\subsection{$T<T_g$}

\begin{figure}[tbp]
\centerline{\epsfxsize=\hsize \epsfbox{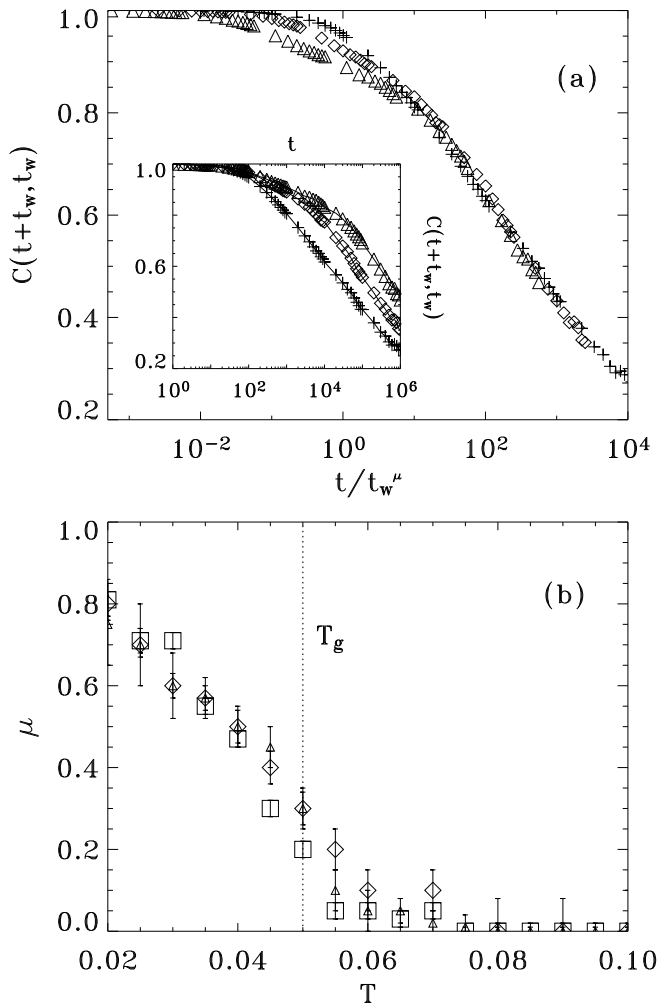}}
\caption{Characterization of aging for $T<T_g$. (a) Autocorrelation
functions $C(t+t_w, t_w)$ {\it vs.} the rescaled time variable 
$t/t_w^{\mu}$ at $T=0.030 < T_g$. The waiting times are $t_w=10^3$ ($+$),
 $t_w=10^4$ ($\Diamond$) and 
$t_w=10^5$ ($\triangle$). The inset shows the same data  
represented as a function of time. (b) Temperature dependence of the aging exponent $\mu$ for different system sizes $N=L^3$ with $L=10$ ($\square$), $L=8$ ($\triangle$) and $L=6$ ($\Diamond$).}
\label{fig:fig3}
\end{figure}

This is the region in which
 non-equilibrium slow relaxation and aging are observed. 
Aging effects can be quantified by performing a non-stationary
 scaling analysis of the two-time 
autocorrelation functions. Experimental
 data in
glasses are often analyzed in terms of the scaling 
form~\cite{review:glasses1,review:glasses2} 
\begin{equation}
C(t + t_w,t_w) \approx F\biggl(\frac{h(t+t_w)}{h(t_w)}\biggr)\;,
\end{equation}
where  $h(u)$  is known as the time-reparameterization function. A
commonly used form  is  $h(u)=\exp(u^{1-\mu}/(1-\mu))$. 
Since this 
form implies an effective time scale growing with $t_w$ as 
$\sim t_w^\mu$ 
we shall analyze our data in terms of the simpler expression  
\begin{equation}
C(t + t_w,t_w) \approx F(t/t_w^{\mu})\;.
\end{equation}
Figure \ref{fig:fig3}(a) illustrates the procedure for $T=0.03$. 
The inset to the figure shows  
$C(t + t_w,t_w)$ as a function of $t$ for three waiting times,  
$t_w=10^3, 10^4$ and $10^5$. The main figure 
presents the same data plotted as a function of $t/t_w^\mu$ with
 $\mu \sim 0.7$, the value for which the collapse of the data
 at large times $t$ is the best. Repeating this procedure for several 
different temperatures we obtained the $T$-dependence of the aging
exponent shown in Fig.~\ref{fig:fig3}(b). We find sub-aging behavior at
low temperatures ($\mu \le 1$) in the range  $T < T_g$. When $T \to
T_g$ $\mu$ decreases steeply to a value close to zero meaning that
aging effects become negligible for $T>T_g \sim 0.05$.  The figure also
shows the size dependence of the aging exponent. It can also be seen that
when the linear size of the system increases from $L=6$ to $L=10$ the
decay of  $\mu$ near $T_g\sim T_c$ becomes steeper. 
This makes it plausible
that $\mu$ vanishes (and aging stops) right at the glass transition
temperature, $T_c$. Confirmation of this hypothesis would require a more
detailed analysis of the $L$ dependence of our results.  

\section{Electron diffusion} 
\label{current}
 The local charge correlation function discussed in the previous
 Section does not provide direct information on the dynamics of current
fluctuations in the medium. Information on this essential  aspect of
the physics  of the Coulomb glass may be obtained from an analysis of
carrier diffusion.

 Let ${\bf r}_i(t)=(x_i(t), y_i(t), z_i(t))$ denote 
the position vector of an electron at time $t$, where $x_i, y_i $ and $z_i(t)$
 are its coordinates {\it before} they are folded
 back into the simulation cell. 
We then have 
\begin{equation}
{\bf r}_i(t)={\bf r}_i(0) + \sum_{k=0}^{N_i(t)} 
\delta {\bf r}_i(k)\;,
\end{equation}
where ${\bf r}_i(0)$ is the electron's initial position, $N_i(t)$ is the 
total number of hops that it performed up to time $t$ and 
$\delta {\bf r}_i(k)$ is the displacement associated with the $k$-th accepted
move.

The mean squared displacement  between times $t_w$ and $t+t_w$ 
 is defined as 
\begin{equation}
\Delta (t + t_w,t_w) = {1 \over M} \sum_{i=1}^M \langle 
\Delta x_i^2 (t + t_w,t_w) 
\rangle\;,
\label{Delta}
\end{equation}
where   
\begin{equation}
\Delta x_i^2 (t + t_w,t_w) = 
\frac{[{\bf r}_i(t+ t_w) - {\bf r}_i(t_w)]^2}{3}\;,
\label{Deltaxi}
\end{equation}
and the angular brackets denote as before an average over realizations
of the disorder, initial conditions and the thermal histories.

\begin{figure}[tbp]
\centerline{\epsfxsize=\hsize \epsfbox{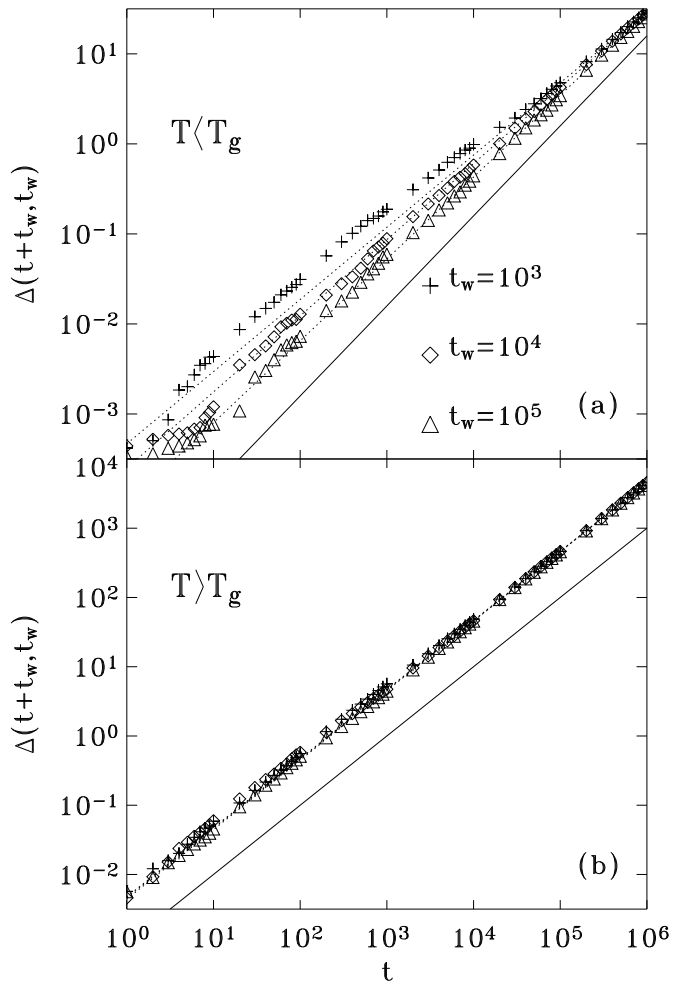}}
\caption{
Mean squared displacement 
$\Delta(t+t_w, t_w)$ as a function of $t$   
for $t_w=10^3$ ($+$), $t_w=10^4$ ($\Diamond$) 
and $t_w=10^5$ ($\triangle$). The solid line is the diffusion limit, $\Delta \propto t$. (a): $T=0.03 < T_g$. The dotted lines are fits of the data to Eq.~(\ref{fitDelta}) over the last decade, $10^5<t<10^6$. (b): $T=0.06 > T_g$. 
}
\label{fig:fig4}
\end{figure}

Figure \ref{fig:fig4} shows the $t$ dependence of
 $\Delta(t+t_w,t_w)$ for three values of the waiting time,
 $t_w=10^3, 10^4$ and $10^5$. Data are displayed 
for two temperatures, $T=0.03<T_g$, [Fig.\ref{fig:fig4}(a)], and
$T=0.06>T_g$, [Fig.\ref{fig:fig4}(b)]. 

It can be seen that $\Delta(t+t_w,t_w)$ is time
translational invariant for the higher temperature but exhibits aging
for the lower one.
Moreover, we observe that in the equilibrium regime $T>T_g$ the average
 motion is diffusive,  $\Delta (t + t_w,t_w)\equiv \Delta (t) \propto
t$, while this is {\it not} the case in the aging regime $T<T_g$. 

\begin{figure}[tbp]
\centerline{\epsfxsize=\hsize \epsfbox{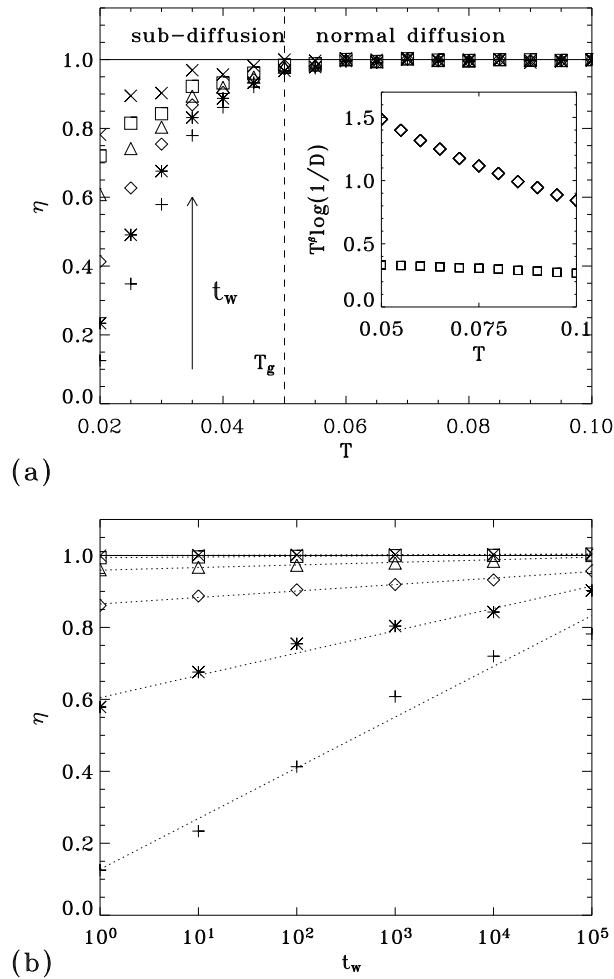}}
\caption{ (a) $T$ dependence of the diffusion exponent $\eta$
 for $t_w = 10^0$ (+), 
$t_w = 10^1$ (*), $t_w = 10^2$ ($\Diamond$), 
$t_w = 10^3$ ($\triangle$), $t_w = 10^4$ 
($\square$) and $t_w = 10^5$ ($\times$), from bottom to top. 
The dashed line indicates the crossover temperature $T_g$. Inset: 
$T$ dependence of the diffusion constant $D$ in the equilibrium 
regime, $T>T_g$. We plot $T^\beta \log D^{-1}$ as a function of $T$ for
  $\beta = 1$ ($\square$) and  $\beta = 0.5$ ($\Diamond$).  
 (b) Waiting-time dependence of $\eta$ 
for $T=0.02$ (+), $T=0.03$ (*), $T=0.04$ ($\Diamond$), 
$T=0.05$ ($\triangle$), $T=0.06$ ($\square$), and $T=0.07$ ($\times$).}
\label{fig:fig5}
\end{figure}

We discuss first the case of of high temperatures. 
In this case we can define a diffusion constant through  
$\Delta (t + t_w,t_w) = D(T)t$.  We attempted 
to fit our results using the stretched-exponential expression 
 $D(T)\sim \exp[-(T_1/T)^{\beta}]$. Using this form, a plot of
 $T^\beta \ln D^{-1}$ as a function of $T$ should result in a horizontal line. 
The inset to  Fig.~\ref{fig:fig5}(a) shows that this is indeed obtained for 
 $\beta \sim 1$. The figure also shows the result of a similar plot using 
the value $\beta=0.5$ expected from the Efros-Shklovski 
variable range hopping law. It is apparent that our data can not be
 described with this value of $\beta$.  Similar deviations from the 
 Efros-Shklovski law were also found in the 2D version of the
 model~\cite{tsigankov,inprog} for which we found $\beta\approx 3/4$.
 \cite{inprog}

We now turn to the analysis of the low-temperature results. 
In this case the time dependence of $\Delta$ can not
 be described by a simple power law but   
we can still characterize the diffusion in this regime 
 by fitting the $t$ dependence of $\Delta (t + t_w,t_w)$ for our
longest times (the last decade in $t$, for example) to an expression of
the form  
\begin{equation}
\Delta (t + t_w,t_w) \sim t^{\eta(t_w,T)}\;.
\label{fitDelta}
\end{equation}

Equation~(\ref{fitDelta})
defines an effective diffusion exponent $\eta$ that depends on both the 
waiting time and the temperature.  
The asymptotic fits for $T=0.03$ 
are represented by the dotted lines in Fig.~\ref{fig:fig4}a   
where the normal diffusion limit, $\eta=1$, is also shown for comparison.

 Figure \ref{fig:fig5} summarizes our results for the diffusion
exponent in the aging region.
The temperature dependence of $\eta$ for several values of $t_w$ is shown
in  Fig.~\ref{fig:fig5}a. It can be seen 
 that in the equilibrium regime, $T>T_g$, 
 $\eta =1$ for all $t_w$ and $T$. Below $T_g$, however,
 the diffusion exponent decreases with decreasing temperature
 for all values of $t_w$ reflecting that electron motion
 becomes increasingly sluggish.

 The dependence of $\eta$ with $t_w$ at fixed $T$ is
 shown in Fig.~\ref{fig:fig5}b for several temperatures. 
The exponent $\eta$
 increases with $t_w$, eventually reaching the diffusion limit $\eta=1$ for
 long waiting times. 
 At low temperature this variation is slow (approximately logarithmic).
This is yet another manifestation of the slow relaxation that
characterizes the glassy phase of the Coulomb glass. 

The characterization of diffusion through a 
diffusion exponent is familiar in the study of 
random walks in random media were one
 generally finds sub-diffusive behavior  
($\eta < 1$) in those physical situations in 
which the distribution of the time intervals between successive hops of
the diffusing particle  has sufficiently long tails that  the  central
limit theorem no longer holds. \cite{review:diffusion} It would be
interesting to examine the distribution of this times in the aging
regime of our system.

\section{Heterogeneous dynamics}
\label{hetero}
To establish a relationship between the observed aging effects
and the microscopic motion of electrons we analyzed
 the evolution of the diffusion front. This latter is defined 
through the probability density of the squared displacements:
\begin{equation}
H_1 (\Delta x^2, t ,t_w) =\frac{1}{M} \sum_i
 \biggl\langle 
\delta 
\bigl(
\Delta x^2 -\Delta x_i^2(t+t_w,t) 
\bigr)
\biggr\rangle\;,
\label{H_1}
\end{equation}
where $\delta(x)$ is the usual $\delta$-function.
It is easy to show that for a stationary and homogeneous  
diffusion process the above distribution takes the form 
\begin{equation}
H_1^{\rm diff} (\Delta x^2, t) =\Phi\left(\frac{\Delta x^2}{D t}-1
\right)\;,
\label{H_1-diff}
\end{equation}
where $\Phi(x)$ is approximately Gaussian. $ H_1^{\rm diff}$ thus exhibits 
 a {\it single} peak whose position increases linearly with time.
\begin{figure}[tbp]
\centerline{\epsfxsize=\hsize \epsfbox{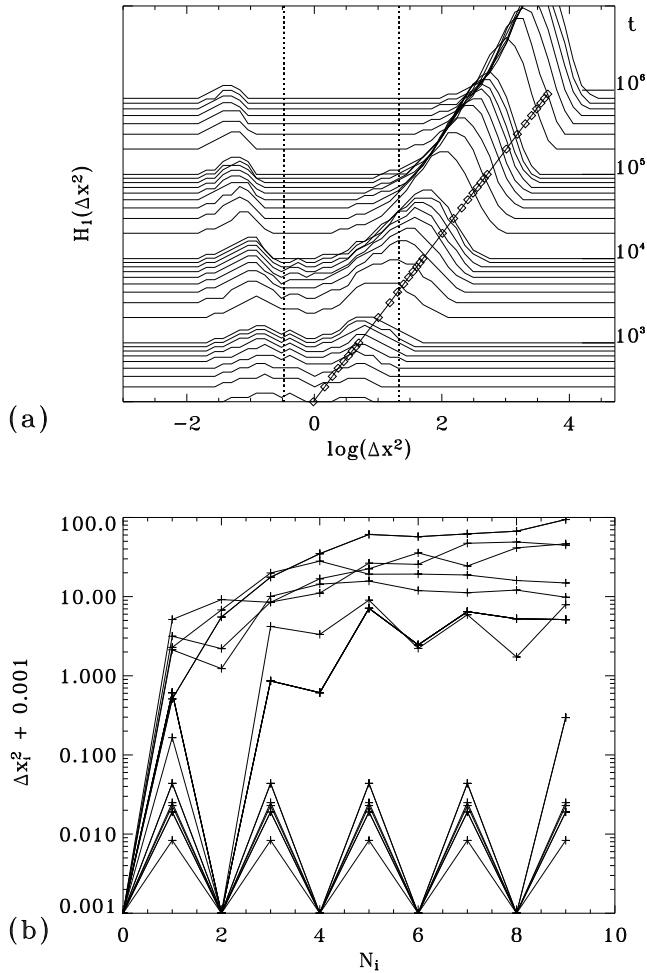}}
\caption{(a) Evolution of the diffusion front $H_1(\Delta x^2)$, for 
$t_w=10^6$ and $T=0.060$. Histograms are scaled and 
shifted vertically by an amount $\log(t)$ so that
 their baselines coincide with the time they correspond to. 
The symbols
($\Diamond$) 
represent the mean squared displacement $\Delta (t + t_w,t_w)$. The vertical 
lines are located at the positions 
$a_0^2$ and $L^2$, where $a_0$ is the 
mean distance between sites and $L=10$ the linear size of the system.
(b) Typical 
squared displacements of a selected set of electrons as a 
function of hop number. The temperature is $T=0.060$. We show 
the ten first hops after $t_w=10^6$ steps.
The selected graphs illustrate typical dipolar and diffusive electron 
trajectories.}
\label{fig:fig6}
\end{figure}
To compute $H_1$ from our numerical data we must 
appropriately coarse-grain the variable $\Delta x^2$. 
Since we found that the distribution of electron 
squared displacements is very broad we chose
a regular coarse graining in $\log(\Delta x^2)$. 
In Fig.~\ref{fig:fig6}(a) we show 
the evolution of the computed $H_1(\Delta x^2,t,t_w)$ as a function of 
$\log(\Delta x^2)$ for $t_w=10^6$ and $T=0.060 > T_g$. For this value
of the temperature $H_1$ is independent  of the waiting time if $t_w$ 
is large enough. This is consistent with the observed time 
translation invariance of the charge autocorrelation functions 
and the mean squared displacement at this temperature. 
The histograms shown in the figure were 
scaled conveniently and shifted vertically by an amount $\log(t)$ to make
 their baselines coincide with the time they correspond to. 

Note that the diffusion front, 
located initially at $\Delta x^2 = 0$, 
splits rapidly in two peaks. The position of the first peak
 is almost time independent at long times. This peak corresponds
to squared displacements 
smaller than the average impurity 
distance, $a_0$. The center of the second peak increases
 linearly with time and its location 
coincides with 
the mean squared displacements at long times.
The interpretation of these results is that the electron dynamics is  
 heterogeneous and characterized by the existence of two dynamical
modes that can be clearly distinguished:  
(a) Diffusive mode: electron
motion is unbound and diffusive. It corresponds 
 to metallic hopping as found in Ref.\onlinecite{lee}; 
(b) Confined mode: electrons remain confined within
 small regions of space during the observation time. 

Examples of trajectories of electrons of these two types are shown in 
Fig.~\ref{fig:fig6}(b) at 
the same temperature and for the same waiting time as above. 
The displacements are represented as a function 
of the hop number $N_i$. We only show 
the ten first hops after a waiting time $t_w=10^6$. 

Some of the trajectories correspond to electrons that hop back and
forth between two sites. $\Delta x^2$ then oscillates between $0$ and
$b_i^2$, the distance  between the sites involved in the motion. 
This fluctuating dipolar motion contributes to 
 most of the weight of the first peak in the histograms of 
Fig.~\ref{fig:fig6}(a). 
It is important to note that, although this fluctuating motion appears 
 regular when plotted as a function $N_i$, it is in fact extremely
 irregular when viewed as a function of  
 time, since the time intervals between successive jumps are very
widely distributed.  
The rest  of the trajectories shown in Figure~\ref{fig:fig6}(b) are those of 
 diffusive electrons that contribute to 
 the metallic peak in the histograms of  Fig. \ref{fig:fig6}(a).

We turn now to the analysis of the statistics of hopping rates which is
 important to understand the source of spontaneous fluctuations in the
 system. To this end we consider the joint distribution function 
\begin{eqnarray}
& H_2 (\Delta x^2, n_h; t ,t_w)=  & \label{H_2}\\
&\sum_i \biggl\langle 
\delta \bigl(\Delta x_i^2 (t + t_w,t_w) - \Delta x^2 \bigr)
\delta \bigl(n_i(t,t_w)-n_h) 
\biggr\rangle & \nonumber
\end{eqnarray}
where $n_i=N_i/\overline{N}_h$ is the number of hops of an electron $i$
normalized by  $\overline{N}_h=\sum_i N_i/ M$, the average number of
hops per electron.

\begin{figure}[tbp]
\centerline{\epsfxsize=\hsize \epsfbox{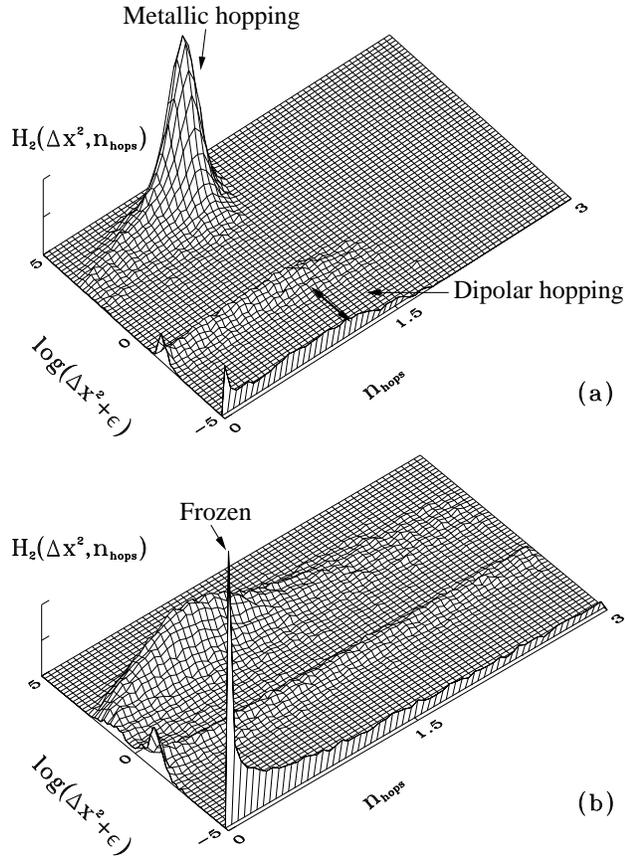}}
\caption{The joint probability distribution of squared 
displacement $\Delta x^2$ and normalized hopping rates  
 $H_2(\Delta x^2, n_h)$ for $T=0.07$ (a) and $T=0.04$ (b).
 The histograms are taken at time $t=10^6$,  
after a waiting time $t_w=10^6$. }
\label{fig:fig7}
\end{figure}

We use a regular coarse graining in $\log(\Delta x^2)$ and $n_h$ 
to compute the 
corresponding histograms. These are displayed in
Figs.~\ref{fig:fig7}(a) and \ref{fig:fig7}(b) for $T=0.070>T_g$ and
$T=0.040<T_g$, respectively. Both plots correspond to $t=t_w=10^6$. 

The two dynamical modes described above can be easily identified in the
equilibrium situation [Fig.~\ref{fig:fig7}(a)]. The single peak at
large $\Delta x^2$ corresponds to the diffusive mode while the two
ridges at low values of $\Delta x^2$ correspond to the dipolar mode.
Note that the hopping rates of electrons involved in dipolar motion
have a much wider distribution than those of diffusive electrons. 

Fig.\ref{fig:fig7}(b) shows   
that the two modes are still distinguishable
at low temperatures, when the system is out of equilibrium. 
The structure of 
the modes is qualitatively different, however. 
Not only the  distribution of hopping rates is now much broader but    
 a small fraction of electrons with low hopping rate   
lies in between the two modes. 
The presence of these electrons, 
that can not be clearly associated with any of the modes, 
suggests that a very slow exchange of 
carriers between them may take place in the course of the relaxation.
We shall further  discuss this issue below.   

To explore in more detail the properties of electrons contributing to
 each of these modes we found it convenient to define for each electron 
 the variable
\begin{equation}
d_i(t_w) = \frac{1}{t} \sum_{\tau=1}^{t} \frac{\Delta x_i^2
 (\tau + t_w,t_w)}{\tau}\;.
\end{equation}
This quantity characterizes the mobility 
of the electron during a time interval of length $t$ after $t_w$. 
For a carrier that diffuses normally in this time span 
with a diffusion constant $D$, $d_i \sim D$. For an electron 
that remained confined in a region of linear size $a$ during 
this time interval, $d_i \sim a^2 \ln(t)/t$. Finally, for a 
frozen electron, {\it i.e.}, one that did not move at all  in the time
interval under consideration, $d_i=0$.

We study the probability density of $d$ defined as
\begin{equation}
f(d,t_w) =\frac{1}{M} \sum_{i=1}^{M}  
\biggl\langle 
\delta \bigl(d_i - d \bigr)
\biggr\rangle\;.
\label{f}
\end{equation}

As discussed above we expect $f(d,t_w)$ to exhibit to well
separated peaks:  one at $d=D(t_w)$, the average diffusion constant of
the diffusing  electrons at time scale $t_w$, and the other at
$d=a(t_w)^2 \ln(t)/t$ where  $a(t_w)$ is the typical size of the
regions of confined motion at the same time scale. The width of the
peaks give the dispersion of these quantities. 

We also introduce the cumulative distribution function
\begin{equation}
F(d,t_w) = \int_{0}^d f(x,t_w) dx\;.
\label{cumm}
\end{equation}  
\begin{figure}[tbp]
\centerline{\epsfxsize=\hsize \epsfbox{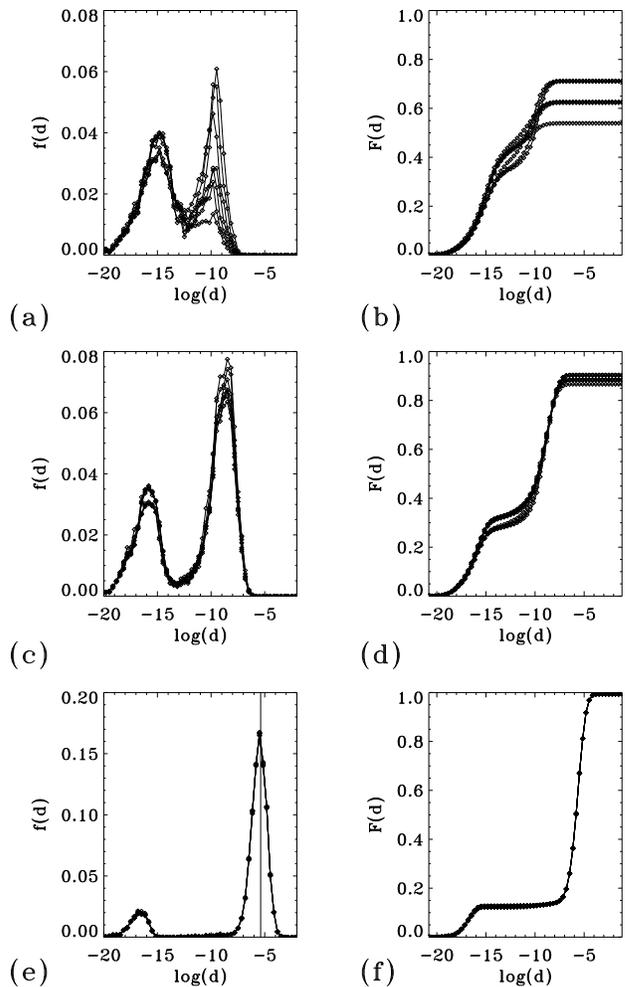}}
\caption{Probability density $f(d)$ (left panels) 
and the corresponding distribution function $F(d)$ (right panels) 
for three temperatures, $T=0.02$ (a,b),  $T=0.035$ (c,d) and  $T=0.06$ (e,f). 
In all cases $t=10^6$ and the waiting times are
 $t_w=10^n$ with $n=0,1,2,3,4,5,6$. The solid line in (e)
  indicates the location of the diffusion constant $D(T)$ 
calculated from the mean squared displacements of electrons at $T=0.060$.}
\label{fig:fig8}
\end{figure}

Results are displayed in Fig.~\ref{fig:fig8} where we plot $f(d, t_w)$,
and its cumulative distribution function  for three temperatures,
$T=0.02 < T_g$, $T=0.035 < T_g$ and $T=0.060>T_g$ and waiting times 
$t_w=10^n$ with  $n=$0,1,2,3,4,5 and 6. In all cases we see the
appearance of the two peaks referred to above. The distributions are
stationary for $T>T_g$ for which the system  equilibrates rapidly but
they show aging in the non-equilibrium regime. 

A striking feature of the function $f$ in the aging regime  [cf.
Fig.~\ref{fig:fig8}(a,c)] is that  the {\it positions} of the peaks are
almost independent of $t_w$. This means that the diffusion constant of
the ``metallic'' electrons is time-independent.  The {\it height} of the
peaks, however, does depend on time scale:  as time elapses from the
quench, the proportion of diffusing carriers diminishes while that of
confined  ones increases. This is a direct manifestation of the
exchange mechanism that we hinted at above. The fact that the  location
of the peaks is only weakly dependent on $t_w$ indicates that  the
effective mobility of the diffusive electrons is not much affected by 
the slow changes of the environment that result from the aging process.

The plateaus that appear  in the  corresponding cumulative 
 distribution functions right after the peaks [cf. Figs.~\ref{fig:fig8}(b,d,f)] 
can be used to measure the relative populations of the modes. 
We see a first plateau corresponding to the area of $f(d,t_w)$ 
below the dipolar peak and a 
second plateau corresponding to the additional area below 
the metallic peak.  
It can be seen that the cumulative distribution function does not 
saturate to unity at low temperatures [cf.  Figs.~\ref{fig:fig8}(b,d)] 
while it does at high temperatures [cf. Fig. \ref{fig:fig8}(f)]. 
The difference is due to the fact that, at low temperature, a fraction
 of the electrons  remain 
frozen during the observation time. These
 were not counted in the numerical evaluation of the integral in 
Eq.~(\ref{cumm}). Another manifestation of the presence of 
 frozen carriers is the pronounced asymmetry of the two
 lower ridges in the lower panel of  
 Fig.~\ref{fig:fig7} in the zero hopping-rate limit. 

\begin{figure}[tbp] 
\centerline{\epsfxsize=\hsize \epsfbox{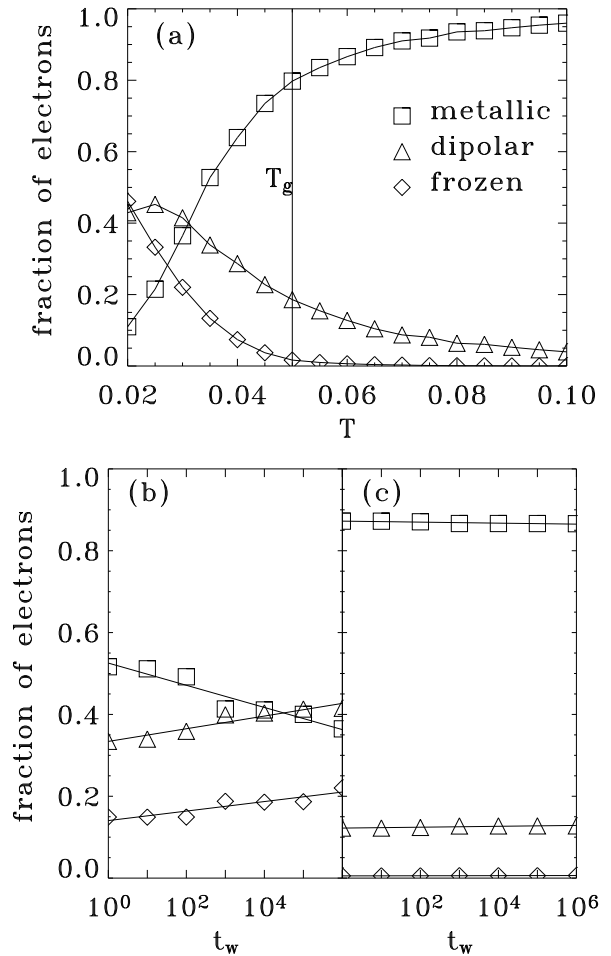}}
\caption{(a) Density of diffusive ($\square$), confined  
 ($\triangle$) and frozen 
electrons ($\Diamond$) during timescale $t=10^6$  
as a function of the temperature. Waiting time is $t_w=10^5$. 
The line indicates the 
location of the crossover temperature $T_g$. 
(b,c) Waiting-time dependence of the same densities  
for $T=0.03<T_g$ (b) and $T=0.06>T_g$ (c). 
The meaning of the symbols is the same as in (a)}
\label{fig:fig9}
\end{figure}

We can now use the height of the plateaus in 
Fig.~\ref{fig:fig8}(b,d,f) to measure the populations of the different modes. 
These are represented as a function of temperature for $t_w=10^6$ in
  Fig.~\ref{fig:fig9}(a). It is seen that the proportion of diffusive 
electrons increases with increasing temperature while, at the same time, that
 of dipolar and frozen ones decreases. Note that a crossover between the regime dominated by diffusive electrons and that dominated by confined ones is located precisely at $T_g$. 

The waiting-time dependence of the populations is shown in 
Fig.~\ref{fig:fig9}(b) for two temperatures, $T=0.03<T_g$ (left panel) and 
$T=0.06>T_g$ (right panel). These populations are time
 independent at the highest temperature but vary logarithmically 
with time in the aging regime. 

It is quite tempting to try to relate these observations to
 the relaxational properties of the conductivity. Assuming that
 the Einstein relation holds in the non-equilibrium regime,
  the conductivity at scale $t$ 
is $\sigma \propto n \int dD f(D,t)$, 
where $n$ is the electron density and the integral extends over the
diffusive peak. We saw earlier that  the position of the peak
$\overline{D}$ is time independent. This implies $\sigma \sim n p_D
\overline{D}$  where $p_D$ is the fraction of diffusing carriers. Since
the latter decreases logarithmically with time, this simple argument
predicts logarithmic relaxation of the conductivity which is one of the
main experimental observations. Whether the assumptions leading to this
result are valid has to await direct computation of the current in the
aging regime, in the presence of an applied  electric field.
\cite{inprog}

\section{Conclusions} 
\label{conclusions} 
We have studied the
relaxational properties of the three dimensional random-site  Coulomb
glass model after a quench from high temperature.  We found a crossover
from stationary to slow  non-stationary dynamics at a temperature $T_g$ that
is very close to $T_c$, the equilibrium glass transition  temperature
of the model. This crossover can be seen even in relatively  small
samples because of the  exponential increase of the equilibration time
with decreasing temperature. 

We found that at low temperature the dynamics of local charge
fluctuations and that of current fluctuations show aging.  In the
former case, the relaxation obeys simple scaling laws characterized 
by a temperature-dependent aging exponent $\mu(T)$. Analysis of 
the temperature and
system-size dependence of $\mu(T)$ suggests that in the thermodynamic
limit  the observed crossover at $T_g\sim T_c$ 
becomes a real dynamic transition  that
occurs precisely at $T_c$.

The analysis of the properties of diffusion fronts revealed that the 
dynamics of carriers is heterogeneous as it was found in other glassy
systems. \cite{dynhet} We found that, for each timescale  two classes
of electrons may be identified,  those that have diffusive motion
during the observation time and those whose motion  in the
same time interval remains confined. 
Only electrons belonging to the former class contribute 
to the dc conductivity
while the others only contribute to the dielectric screening.   

In the region of low temperatures where aging is observed   electrons
are slowly exchanged between these two modes with the consequence that
the population of metallic electrons decreases logarithmically with
time without appreciable change of the their diffusion constant. This
provides a plausible explanation for the logarithmic relaxation of the
conductance after a quench that was observed experimentally. 

We believe that the local microscopic dynamics used here, which is a 
realistic description of hopping processes in experimental systems,
plays an important role in the phenomena that we observed.  
This type of dynamics favors the appearance of 
local effective constraints that relate our model to  
kinetically constrained models in which slow dynamics 
arises from restrictions on the allowed transitions between 
configurations.

\acknowledgments

We thank Z.Ovadyahu, L. F. Cugliandolo, J. Kurchan, and 
T. Giamarchi for valuable discussions. 

We acknowledge finantial support from the Argentina-Francia 
cooperation SECYT-ECOS, project A01E01. A.B.K. and D.D.
aknowledge support from Conicet, CNEA, 
ANPCyT (PICT99-03-06343) and Fundaci\'{o}n Antorchas
(Proy. 14116-147). A.B.K. also aknowledges support from 
a grant of the Swiss National Science Foundation.

\newpage


\begin{thebibliography}{}

\bibitem{films:experiments} M. Ben-Chorin, Z. Ovadyahu and
M. Pollak, Phys. Rev. B {\bf 48} 15025 (1993); Z. Ovadyahu
and M.Pollak, Phys. Rev. Lett. {\bf 79}, 459 (1997).

\bibitem{films:goldman} G. Martinez-Arizala, C. Christiansen, D. E. Grupp,
N. Markovic, A. M. Mack, and A. M. Goldman,  Phys. Rev. B {\bf 57} 670 (1998).

\bibitem{films:aging}
A. Vaknin, Z. Ovadyahu and M.Pollak, Phys. Rev. Lett. {\bf 84},
3402 (2000); {\it ibid.}, Phys. Rev. B {\bf 65}, 134208 (2002).

\bibitem{films:ee-int}
A. Vaknin, Z. Ovadyahu and M.Pollak, Phys. Rev. Lett. {\bf 81},
669 (1998).

\bibitem{orlyanchik}
V. Orlyanchik and Z. Ovadyahu, Phys. Rev. Lett. {\bf 92},
066801-1 (2004).

\bibitem{review:coulombglass1} B.I.Schklovskii and A.L. Efros, 
{\it Electronic Properties of Doped Semiconductors 
(Springer, Berlin, 1984).}

\bibitem{coulomb-glass} {\it Electron-Electron interactions in Disorder
Systems}, edited by A. L. Efros and M. Pollak (North-Holland,
Amsterdam, 1985). 

\bibitem{pollak1} M. Pollak,
Philos. Mag. B {\bf 50}, 265 (1984); M.Gr\"{u}newald {\it et al.},
J. Phys. C {\bf 15}, L1153 (1982).

\bibitem{davies} J. H. Davies, P. A. Lee, and T. M. Rice,
Phys. Rev. Lett. {\bf 49}, 758 (1982); {\it ibid.}, Phys. Rev. B
{\bf 29}, 4260 (1984).

\bibitem{lee} W. Xue and P. A. Lee, Phys. Rev. B {\bf 38},
9093 (1988).

\bibitem{yu1} E. R. Grannan and Clare C. Yu,
Phys. Rev. Lett. {\bf 71}, 3335 (1993).

\bibitem{newstuff} For recent developements see
M. Mueller and L. B. Ioffe, cond-mat/0406324; 
S. Pankov and V. Dobrosavljevic, cond-mat/0406406.

\bibitem{ortuno}  A. P\'erez-Garrido, M. Ortu\~no, A.
D\'{\i}az-S\'anchez and E. Cuevas, Phys. Rev. B {\bf 59}, 5328 (1999); 

\bibitem{yu2} C. C. Yu,
Phys. Rev. Lett. {\bf 82}, 4074 (1999).
 
\bibitem{tsigankov} D. N. Tsigankov, {\it et al}, Phys. Rev. B {\bf
68}, 184205 (2003). 

\bibitem{grempel} D. R. Grempel, Europhys. Lett. {\bf 66}, 854 (2004).
 
\bibitem{review:glasses1}
L. C. E. Struik, {\it Physical Aging in Amorphous Polymers and
Other Materials} (Elsevier, Amsterdam,1978).

\bibitem{review:glasses2} E. Vincent {\it et
al}, in {\it Complex behavior of Glassy Systems}, M. Rubi and C.
Perez-Vicente Eds. (Springer, Berlin, 1997).

\bibitem{review:glasses3} J.P. Bouchaud, L.F. Cugliandolo, and J. Kurchan in 
{\it Spin Glasses and Random Fields}, Ed.: A.P.Young (World Scientific, 
Singapore, 1997); L. F. Cugliandolo, in {\it Slow Relaxation and non 
equilibrium dynamics in condensed matter}, Les Houches Session 77 July 2002
J-L Barrat, J Dalibard, J Kurchan, M V Feigel'man eds, cond-mat/0210312.  

\bibitem{CuKu} L. F. Cugliandolo and J. Kurchan, Phys. Rev. Lett.
{\bf 71}, 173 (1993); {\it ibid}, J. Phys. A {\bf 27}, 5749 (1994).

\bibitem{dynhet} J. P. Garrahan and D. Chandler, Phys. 
Rev. Lett. {\bf 89}, 035704 (2002);
L. Berthier and J. P. Garrahan, Phys. Rev. E {\bf 68}, 
041201 (2003); S. Whitelam, L. Berthier and J. P. Garrahan, Phys. Rev. Lett. 
{\bf 92}, 185705 (2004).

\bibitem{dynhet2} H. E. Castillo, C. Chamon, L. F. Cugliandolo, 
J. L. Iguain, and M. P. Kennett, Phys. Rev. B {\bf 68}, 134442 (2003).

\bibitem{ritort} F. Ritort and P. Sollich, Adv. Phys. {\bf 52}, 219 (2003).

\bibitem{noise1} B. I. Shklovskii, Phys. Rev. B {\bf 67}, 045201 (2003). 

\bibitem{noise2} K. Shtengel and C. C. Yu, Phys. Rev. B {\bf 67}, 165106 (2003).
 
\bibitem{transiciondeglass} T. Vojta and M. Schreiber, Phys. Rev. Lett.
{\bf 73}, 2933 (1994);  E. R. Grannan and C. C. Yu, Phys. Rev. Lett
{\bf 73}, 2934 (1994);   A. D\'{\i}az-S\'anchez,  M. O. Ortu\~no, A.
P\'erez-Garrido, and  E. Cuevas, Phys. Stat. Sol. (b) {\bf 218}, 11
(2000).

\bibitem{inprog} A.B. Kolton, D. Grempel and D. Dom\'{\i}nguez, unpublished.  


\bibitem{review:diffusion} J.P. Bouchaud, A. Georges, Physics Reports 
{\bf 195}, 127-293, (1990).

  
\end{thebibliography}
\end{document}